\documentclass{ws-p9-75x6-50}

\begin{document}

\title{Signals of primordial phase transitions on CMB maps}
\author{P.S. Corasaniti$^{1,2}$, L. Amendola$^{2}$, F. Occhionero$^{2}$}

\address{$^{1}$Centre for Theoretical Physics, University of Sussex, Falmer,
 Brighton BN1 9QJ, United Kingdom}

\address{$^{2}$Osservatorio Astronomico di Roma, Viale del Parco Mellini 84, 00136
Roma, \\Italy}

\maketitle
\abstracts{
The analysis of the CMB anisotropies is a rich source of cosmological informations.
In our study, we simulated the signals produced by the relics of a first order phase transition
occured during an inflationary epoch in the early Universe. These relics are bubbles of true 
vacuum that leave a characteristic non-Gaussian imprint on the CMB. We use different 
statistical estimators in order to evaluate this non-Gaussianity. We obtain some limits 
on the allowed values of the bubble parameters comparing our results with the experimental data. We also predict the possibility to detect this signal with the next high resolution experiments.}

\section{Introduction}

The models of structure formation assume the existence of primordial matter
density perturbations, that at the decoupling are source of the temperature
fluctuations in the Cosmic Microwave Background (CMB). Currently we have two
sets of theories that can explain the formation of these initial
perturbations. In the inflationary models with second order phase
transition, the quantum fluctuations of the Inflaton field produce Gaussian
and scale invariant density perturbations. On the other hand, in models with
a transition of the first order, as topological defects and extended
inflation, the perturbations are non-Gaussian and scale dependent. Therefore
the statistical properties of the CMB anisotropies are a powerful tool in
order to distinguish these models. In our study we analyse the
non-Gaussianity produced by the relics of a first order phase transition in the
context of a specific realization of the extended inflation, built on a
non-minimal generalization of Quadratic Gravity$^{1}$. These relics are
bubbles of true vacuum that contribute together with the ordinary quantum
perturbations to structure formation. In particular these spherical defects
can explain the presence of voids observed in several galaxy surveys$^{2}$.
The parameters describing the void distribution are the central density
contrast $\delta $ of the bubble, its radius $R$ and the fraction $X$ of the
space filled by the voids. At the decoupling the bubbles leave a
characteristic imprint on the CMB$^{3}$. In fact solving the linearized 
Boltzmann-Einstein equation for a bubble perturbation, intersecting the last
scattering surface, we have a signal that appear like a cold spot surrounded by
an hot isothermal ring, on the scale of the sound horizon at the decoupling.
Therefore the bubble imprint is a source of non-Gaussianity.

\section{Non-Gaussian analysis}

We studied the statistical properties of the bubble anisotropies with two 
different estimators, the three point collapsed function $%
C_{3}(\alpha )$ and the normalised bispectrum$^{4}$ $I_{l}^{3}$. Both
estimate the third order correlation of the CMB\ anisotropies, the former in
the real space and the latter in the multipole space. \\The $C_{3}(\alpha )$ has been used 
to compare the bubble non-Gaussianity with the cosmic variance of low resolution 
observations (COBE). Using the formalism of
Magueijo$^{5}$ for the texture-spot anisotropies we 
calculated an analytical
formula of the 3-collapsed function for a possonian 
bubble distribution on
the sky. We considered voids with a present radius of $30$ 
$h^{-1}Mpc$ and with a filled space fraction $X\sim 0.3,0.5$, in
agreement with the properties of the voids observed in the galaxy surveys. With 
these parameters the three collapsed function has been evaluated for different values
of the bubble density contrast.
Comparing the results with the COBE 4 YEARS data
$^{6}$, we obtain an upper limit on $\delta$ $^{7}$. In particular from a chi-square
analysis we have $\delta \leq 0.0017$ and $X\leq 0.54$. \\ In high resolution
experiments the bubble signal should be more evident. In order to detect it with a
non-Gaussian analysis we used the $I_{l}^{3}$ bispectrum. This estimator is very 
efficient and not to depend on the cosmological model, being normalized with an 
appropriate power of the CMB spectrum. We inferred the probability distribution function of the
bispectrum, in the range of multipoles $l\in
(150,250)$, by Monte Carlo simulations of full CMB
sky maps with a resolution of $10^{\prime }$ arcmin. We have that the distributions obtained by the Gaussian
simulations are different from those obtained by bubble ones. In particular the
latter have a variance much larger than in the Gaussian case. On the
contrary at low multipoles these probability functions become identical, as we
expect from the Central Limit Theorem. Therefore, the next high resolution
observations from Map and Planck could detect and distinguish 
the non-Gaussianity produced by primordial voids.

\end{document}